\documentclass[aps,prd,onecolumn,superscriptaddress]{revtex4}
\usepackage{epsfig}
\usepackage{epstopdf}
\input{epsf.sty}
\usepackage{epsf}
\usepackage[hypertex]{hyperref}
\usepackage{natbib,ifthen}
\usepackage{color}

\newcommand{\adsurl}[1]{\href{#1}{ADS}}
\usepackage{amssym,aas_macros}

\newcommand\lsim{\mathrel{\rlap{\lower4pt\hbox{\hskip1pt$\sim$}}
        \raise1pt\hbox{$<$}}}
\newcommand\gsim{\mathrel{\rlap{\lower4pt\hbox{\hskip1pt$\sim$}}
        \raise1pt\hbox{$>$}}}
\def\beq{\begin{equation}}
\def\eeq{\end{equation}}
\def\bea{\begin{eqnarray}}
\def\eea{\end{eqnarray}}

\def\tr{}

\date{\today}

\def\LaTeX{L\kern-.36em\raise.3ex\hbox{a}\kern-.15em
    T\kern-.1667em\lower.7ex\hbox{E}\kern-.125emX}

\begin{document}
\title{Extended Limber Approximation}
\author{Marilena LoVerde}\email{marilena@phys.columbia.edu}
\affiliation{Institute for Strings, Cosmology and Astro-particle Physics (ISCAP)}
\affiliation{Department of Physics, Columbia University, New York, NY 10027}
\author{Niayesh Afshordi}\email{nafshordi@perimeterinstitute.ca}
\affiliation{Perimeter Institute
for Theoretical Physics, 31 Caroline St. N., Waterloo, ON, N2L 2Y5,Canada}
\begin{abstract}
\tr{
We develop a systematic derivation for the Limber approximation to the angular cross-power spectrum of two random fields, as a series expansion in $(\ell+1/2)^{-1}$. This extended Limber approximation can be used to test the accuracy of the Limber approximation and to improve the rate of convergence at large $\ell$'s. We show that the error in ordinary Limber approximation is ${\cal O}(\ell^{-2})$. We also provide a simple expression for the $2^{nd}$ order correction to the Limber formula, which improves the accuracy to ${\cal O}(\ell^{-4})$. This correction can be especially useful for narrow redshift bins, or samples with small redshift overlap, for which the $0^{th}$ order Limber formula has a large error. We also point out that using $\ell$ instead of $\ell+1/2$, as is often done in the literature, spoils the accuracy of the approximation to ${\cal O}(\ell^{-1})$.
}
\end{abstract}

\maketitle
 \section{Introduction}
Many observations in cosmology are observations of random fields (e.g. the cosmic microwave background (CMB) anisotropy, fluctuations in the mass density or galaxy distribution, the weak lensing shear or convergence field, and 21cm \tr{emission line} fluctuations). A primary means to learn about the distribution and evolution of large-scale structure are through correlation functions of these fields, the simplest being the two point correlation function or its Fourier transform\tr{,} the power spectrum. Many observations are given in terms of the angular correlation function $w_{AB}(\hat{n}\cdot\hat{n}')$ or its spherical harmonic transform\tr{,} the angular power spectrum $C_{AB}(\ell)$
\beq
w_{AB}(\hat{n}\cdot\hat{n}') \tr{\equiv} \langle A(\hat{n})B(\hat{n}')\rangle =\sum_{\ell}\frac{2\ell+1}{4\pi}C_{AB}(\ell)P_\ell(\hat{n}\cdot\hat{n}')
\eeq
where $A$ and $B$ are line-of-sight projections of the fields being correlated (e.g. the temperature anisotropy $\Delta T/T$ or the mass fluctuation $\delta\rho/\rho$), $\hat{n}$, $\hat{n}'$ are unit vectors indicating the direction of observation and $P_{\ell}$ are the Legendre polynomials.

Calculations of angular power spectra give expressions in terms of several integrals which must be evaluated numerically. The Limber approximation \cite{Limber} and its generalization to Fourier space \cite{Kaiser,Kaiser2} is a commonly used technique to simplify calculations. In implementing the Limber approximation one assumes small angular separations (or large multipole moment $\ell$) and that some of the functions being integrated are more slowly varying than others. The Limber approximation is 
\tr{powerful method} to \tr{accurately} estimate the magnitude 
and understand the \tr{analytic} dependencies of the \tr{projected} power spectra. Also, since the Limber approximation reduces the number of integrals numerical calculations are simpler.

In this paper\tr{,} we present a \tr{systematic} derivation of the Limber approximation to the angular power spectrum as a series expansion in $(\ell+\frac{1}{2})^{\tr{-1}}$, \tr{which is a rigorous generalization of a technique introduced in \cite{ALS}}. While the first term in the expansion is the usual Limber approximation, higher order terms can be considered as an extension. We apply this approximation to a few examples where keeping additional terms in the expansion might be desirable. The results presented here can be applied to the cross-correlation of two random fields whose Fourier space power spectr\tr{a are} isotropic. An analysis of the Limber approximation and a proposed alternative approximation for the real space correlation function is given in \cite{Simon}. For another discussion of some issues related to the validity of the Limber approximation for lensing power spectra
\tr{, the reader can refer to} Appendix C of \cite{Dodelson}.

 In \S \ref{LimberDerivation}, we present the derivation of the extended Limber approximation for the angular cross-power spectrum of two random fields. In \S \ref{FlatSky}, we make a comparison with the flat sky approximation. \S \ref{Examples} \tr{applies} the derived first and second terms in the Limber approximation to a few examples: the galaxy auto-power spectrum, the cross-power spectrum of two redshift bins with small overlap in redshift, and the cross-power spectrum of broad and narrow redshift distributions. Concluding remarks are given in \S \ref{Conclusion}.

\section{Derivation of the Limber Approximation}
\label{LimberDerivation}
We first develop the theoretical expectation value of the cross-correlation of two random fields, projected on the sky.  Let us consider two random fields $A({\bf x})$ and $B({\bf x})$ with their Fourier transforms defined as
\beq
A({\bf k}) = \int  d^3{\bf x}~  e^{-i{\bf k\cdot x}} A({\bf x}) ~ {\rm and} ~B({\bf k}) =   \int d^3{\bf x}~  e^{-i{\bf k\cdot x}}B({\bf x}).
 \eeq
These fields could be, for instance, the \tr{density} fluctuation $\delta\rho({\bf x})/\rho$ or the Newtonian potential $\Phi({\bf x})$. The cross-correlation power spectrum, $P_{AB}(k)$ (which is assumed to be isotropic) is defined by
\beq
\langle A({\bf k_1} )B^*({\bf k_2})\rangle = (2\pi)^3 \delta^3({\bf k_1-k_2}) P_{AB}(k_1).
\eeq
The projections of $A$ and $B$ on the sky are defined using $F_A$ and $F_B$ projection kernels
\beq
\tilde{A}({\bf \hat{n}}) \,=\, \int dr~F_A(r) A(r{\bf \hat{n}}),\,~{\rm and}~ \tilde{B}({\bf \hat{n}}) \,=\, \int dr~F_B(r) B(r{\bf \hat{n}}).
\eeq
Now, expanding $\tilde{A}$ and $\tilde{B}$ in terms of spherical harmonics, the angular cross-power spectrum, $C_{AB}(\ell)$ is defined as
\bea
\label{Clexact}
C_{AB}(\ell)&\equiv& \langle\tilde{A}_{\ell m}\tilde{B}^*_{\ell m}\rangle\nonumber\\
&=& \int dr_1 dr_2 F_A(r_1) F_B(r_2) \int \frac{ d^3{\bf k}}{(2\pi)^3} P_{AB}(k) (4\pi)^2 j_{\ell}(kr_1)j_{\ell}(kr_2)Y_{\ell m}({\bf \hat{k}})Y^*_{\ell m}({\bf \hat{k}})\nonumber\\
&=& \int dr_1 dr_2 F_A(r_1) F_B(r_2) \int \frac{2k^2dk}{\pi}j_{\ell}(kr_1)j_{\ell}(kr_2) P_{AB}(k)
\nonumber\\
&=& \int\! k\,dk \,P_{AB}(k) \int dr_1 f_A(r_1)J_{\ell+1/2}(k r_1)  \int dr_2 f_A(r_2)J_{\ell+1/2}(k r_2),
 \eea
 where $j_{\ell}$'s are the spherical Bessel functions of rank $\ell$ and $Y_{\ell m}$'s are the spherical harmonics. In the last step, we have substituted the spherical Bessel functions in terms of the Bessel functions of the first kind, $J_{\ell+1/2}$, and defined:
\beq f_A(r)\equiv \frac{F_A(r)}{\sqrt{r}}; f_B(r) \equiv\frac{F_B(r)}{\sqrt{r}}.
\eeq
At the next step, we will develop a series representation for the integral of an arbitrary function multiplied by the Bessel function. We will use the fact that the Bessel function (for $\nu>0$) grows monotonically from zero at $x=0$ to $x\simeq \nu$ and starts oscillating rapidly afterwards, to write:
 \bea
 \label{SeriesBessel}
 \lim_{\epsilon \rightarrow 0}\int_0^{\infty} e^{-\epsilon (x-\nu)} f(x) J_{\nu}(x) dx = B_0 f(\nu)+B_1 f'(\nu)+ B_2 f''(\nu)+B_3 f'''(\nu) + ...
\eea
Using the Taylor expansion of $f(x)$ around $x=\nu\tr{\equiv}\ell+1/2$, we find:
 \bea
B_n &=&\frac{1}{n!} \lim_{\epsilon \rightarrow 0}\int_0^{\infty} e^{-\epsilon (x-\nu)} (x-\nu)^n J_{\nu}(x)dx \\
 &=& \frac{(-1)^n}{n!} \lim_{\epsilon \rightarrow 0}\frac{\partial^n}{\partial\epsilon^n}\int_0^{\infty} e^{-\epsilon (x-\nu)}  J_{\nu}(x)dx.
\eea
The integral over the Bessel function is a standard Laplace transform, which has \tr{a} closed form:
 \bea
  \label{LaplaceBessel}
\int_0^{\infty} e^{-\epsilon (x-\nu)}  J_{\nu}(x)dx = e^{\epsilon\nu}\frac{\left(\sqrt{1+\epsilon^2}+\epsilon\right)^{-\nu}}{\sqrt{1+\epsilon^{2}}}
 =    1-\frac{\epsilon^2}{2}+\frac{\nu\epsilon^3}{6}+\frac{3\epsilon^4}{8}-\frac{19\nu\epsilon^5}{120} + O(\epsilon^6)\,,\qquad\qquad
\eea
\tr{yielding}:
 \bea
B_0=1, B_1=0, B_2= -\frac{1}{2},B_3=-\frac{\nu}{6},B_4=\frac{3}{8}, B_5=\frac{19\nu}{120},\dots\nonumber
 \eea
Therefore, we find
\bea
C_{AB}(\ell) = \int dk ~k~  P_{AB}(k)\left(k^{-1}f_A(r)-\frac{k^{-3}}{2} f''_A(r)-\frac{\nu k^{-4}}{6}f'''_A(r)+...\right)\nonumber\\
\times \left(k^{-1}f_B(r)-\frac{k^{-3}}{2} f''_B(r)-\frac{\nu k^{-4}}{6}f'''_B(r)+...\right)\,,\qquad\qquad
\eea
where $kr=\nu=\ell+1/2$.  Combining the parentheses and collecting terms of the same order in $\nu$ one finds,
\bea
C_{AB}(\ell) =\int \frac{dr}{r}P_{AB}(\frac{\nu}{r})f_A(r)f_B(r)\left\{1-\frac{1}{\nu^2}\left[\frac{r^2}{2}\left(\frac{f''_A(r)}{f_A(r)}+\frac{f''_B(r)}{f_B(r)}\right)+\frac{r^3}{6}\left(\frac{f_A'''(r)}{f_A(r)}+\frac{f_B'''(r)}{f_B(r)}\right)\right]+\mathcal{O}(\nu^{-4})\right\}
\eea
After algebraic manipulations and some integrations by parts\tr{,} the two parentheses in the $1/\nu^2$ term  can be combined to find:
\bea
 \label{ClexLimber}
 C_{AB}(\ell)= \int \frac{dk}{k} P_{AB}(k) f_A(r)f_B(r) \left\{1+
 \frac{\nu^{-2}}{2}\left[\frac{d\ln f_A}{d\ln r}\frac{d\ln f_B}{d\ln r}s(k)-p(k)\right]+O(\nu^{-4})\right\}\,,
\eea
where
\beq s(k) = \frac{d\ln P_{AB}(k)}{d\ln k}, \, p(k) = \frac{k^2[3P''_{AB}(k)+kP'''_{AB}(k)]}{3P_{AB}(k)}.\label{s+p}
\eeq

\tr{Equations (\ref{ClexLimber}-\ref{s+p}) show the first systematic correction to the Limber approximation, which can be used to reduce the error in the approximation from $\ell^{-2}$ to $\ell^{-4}$. Moreover, we can use relative magnitude of the sub-leading term in the expansion as a criterion for the convergence/reliability of the Limber approximation.}
\tr{We thus see that} the convergence of the Limber expansion depends on both $\nu=\ell+1/2$ and the $f_{A}$, $f_{B}$.  If the two kernels $f_A$ and $f_B$ are peaked at the same distance $\bar{r}$, the $1/\nu^2$ term is subdominant when $\nu \gsim \bar{r}/{\rm max}[\sigma_A,\sigma_B]$ where $\sigma_A$ is the width of $f_A$ and $\sigma_B$ is the width of $f_B$. \tr{However,} if $f_A$ and $f_B$ are peaked at different distances, say $r_A\gsim r_B+\sigma_B$, where $r_A$ and $r_B$ are the locations of the maxima, truncating the expansion requires $\nu \gsim \bar{r}(r_A-r_B)/\sigma_A\sigma_B$ \footnote{\tr{Here, we assumed that $f_A$ and $f_B$ can be approximated as Gaussian.}}.

 \section{Flat Sky and $\ell+1/2$}
 \label{FlatSky}
Let us now think about the 2D power spectrum in the flat sky limit (for a comparison with the angular power spectrum see also \cite{HuFlatSky}). \tr{To do} this\tr{,} we will use cartesian coordinates with $x_{||}$ the line-of-sight direction and ${\bf x_\perp}$ the perpendicular direction, \tr{and integrate \tr{along} the $x_{||}$ direction}
\beq
A({\bf x_\perp}) \,=\, \int dx_{||}~F_A(x_{||}) A({\bf x_\perp},x_{||}),
 \eeq
and
\beq
A({\bf k_\perp}) \,=\int d^2{\bf x_\perp}\, e^{-i{\bf k_\perp}\cdot {\bf x_\perp}}\int dx_{||}~F_A(x_{||}) A({\bf x_\perp},x_{||}),
\eeq
so the 2D power spectrum will be given by
\beq
\langle A({\bf k_\perp})B({\bf k_\perp'})\rangle= \int dx^A_{||} F_A(x^A_{||})\int dx^B_{||}F_B(x^B_{||})\int \frac{dk_{||}}{(2\pi)}e^{ik_{||}(x^A_{||}-x^B_{||})}P_{AB}({\bf k_\perp},k_{||})(2\pi)^2\delta^{(2)}({\bf k_\perp}+{\bf k_\perp'}).
\eeq
Expanding the power spectrum about $k_{||}=0$ (and assuming $P_{AB}({\bf k}_\perp,k_{||})=P_{AB}(\sqrt{k_\perp^2+k_{||}^2})$) gives
\beq
\label{P2exLimber}
P_{2D}(k_\perp)=\int dx_{||}F_A(x_{||})F_B(x_{||})P_{AB}( k_\perp)
\left\{1+\frac{1}{2}\frac{1}{x_{||}^2k_\perp^2}\left(\frac{d\ln P_{AB}}{d \ln k}\frac{d\ln F_A(x_{||})}{d\ln x_{||}}\frac{d\ln F_B(x_{||})}{d\ln x_{||}}\right) +\mathcal{O}\left((k_\perp x_{||})^{-4}\right)\right\},
\eeq
\tr{where the derivatives of $P_{AB}$ are evaluated at $k_{||}=0$.} Notice that all of the $P_{AB}$ factors are independent of $x_{||}$.
How to compare this to the angular power spectrum?  We expect
\beq
 \ell(\ell+1)C_\ell \approx k_\perp^2 P(k_\perp)
\eeq
for large $\ell$.  Expanding $\frac{1}{r^2}P(\nu/r)$ in Equation (\ref{ClexLimber}) about $r=\bar{r}$ where $\bar{r}$ is, for example, the peak distance of $F_{A}(r)F_B(r)$ and comparing with Equation (\ref{P2exLimber})  we can see that indeed $\ell(\ell+1)C_\ell\approx k_\perp^2 P_{2D}(k_\perp)$ for $\ell+1/2=\bar{r}k_\perp$.  Notice that while for large $\ell$\tr{'s}, $\ell+1/2\approx \ell$\tr{,} at \tr{small} $\ell$\tr{'s} the factor of $1/2$ actually makes a difference. Comparing the Laplacian in spherical coordinates with the Laplacian in Fourier space shows that indeed $k\bar{r}\rightarrow \sqrt{\ell(\ell+1)}=\ell+1/2+\mathcal{O}(1/\ell)$ is the correct replacement.

\section{Examples and Comparison of Limber and Exact results at different orders}
\label{Examples}
Here we consider a few examples of calculations of angular power spectra. For simplicity we will assume a spatially flat cosmology so that $r=\chi(z)$ is the comoving distance. In all plots we assume a $\Lambda$CDM universe with $\Omega_m=0.27$, $\Omega_\Lambda=0.73$, $\Omega_b=0.046$ as the fractional densities of matter, cosmological constant and baryons, Hubble constant today \tr{$H_0= 70 ~{\rm km/s/Mpc}$},  scalar fluctuation amplitude $\sigma_8=0.8$,  and scalar spectral index $n_s=0.95$. For the linear matter power spectrum we use the transfer function of \cite{EH98}.

\subsection{Power spectrum of a narrow redshift bin}
\begin{figure}
\begin{tabular}{cc}
\includegraphics[width=0.5\textwidth]{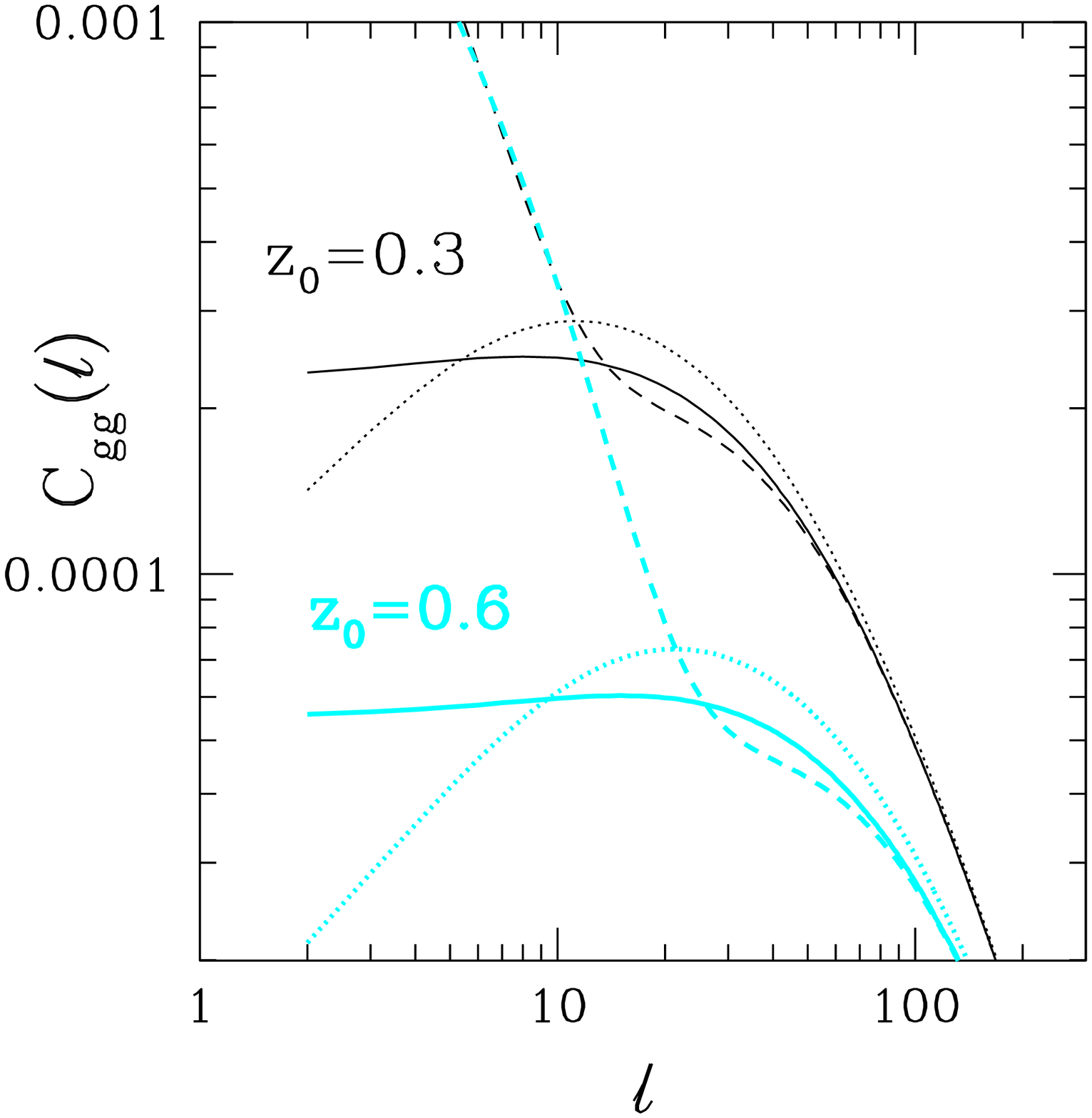}& \includegraphics[width=0.5\textwidth]{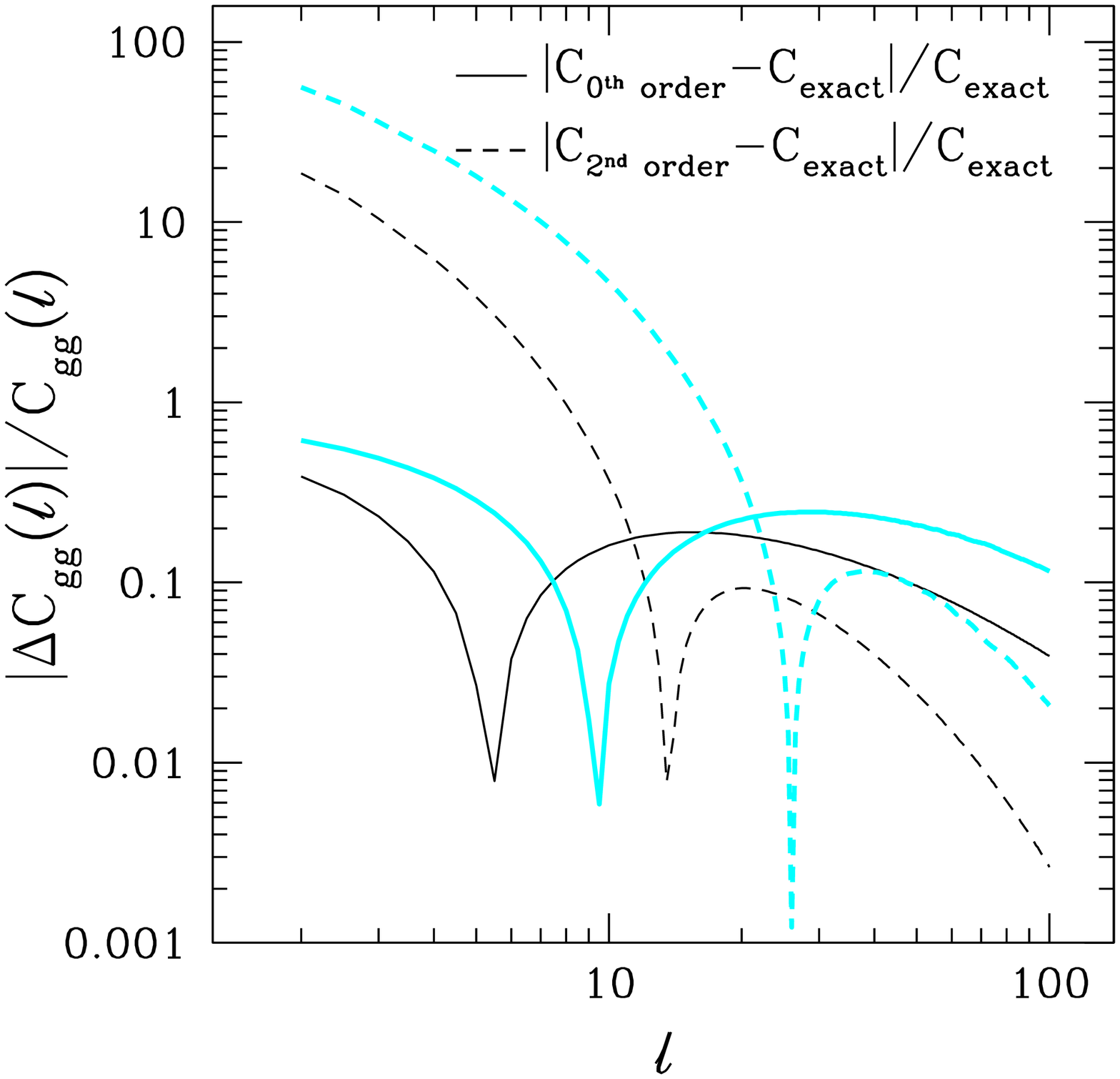}
\end{tabular}
\caption{Left Panel: The galaxy auto-correlation for a narrow redshift bin ($\sigma_{z}=0.01$) at redshifts $z_0=0.3$ (upper black curves) and $z_0=0.6$ (lower cyan/gray curves). The solid lines are the exact $C_{\ell}$ (Equation (\ref{Clexact})) the dotted lines are the $0^{th}$ order Limber approximation, the dashed lines are the Limber approximation keeping the first $\mathcal{O}(\nu^{-2})$ term in Equation (\ref{ClexLimber}). Right panel: The difference between the Limber approximation at $0^{th}$ and $2^{nd}$ order (in $1/\nu$) and the exact angular power spectrum in redshift bins at $z_0=0.3$ (black) and $z_0=0.6$ (cyan/gray). \tr{The radii of convergence for the Limber expansion are roughly at $\ell \sim 15$ and $30$ respectively.}}
\label{NarrowBinCompare}
\end{figure}


From Equation (\ref{ClexLimber}) we can see that keeping the first term in the Limber approximation is accurate only so long as the functions $f_{A}(r)$ and $f_B(r)$ are slowly varying.  Here we calculate the galaxy auto-power spectrum (for instance, \cite{Tegmark,Blake,Frith}). In calculating the galaxy auto-power spectrum (ignoring non-linear evolution), $C_{gg}(\ell)$ these kernels take the form
\beq
\label{littlefg}
f_g(r)= \frac{H(z)}{c} \frac{W(z,z_0)}{\sqrt{r(z)}}D(z)
\eeq
where $H(z)$ is the Hubble parameter, $D(z)$ is the linear growth function with $D(z=0)=1$, $c$ is the speed of light and $W(z,z_0)$ is a normalized selection function centered at $z_0$. If we assume that the linear galaxy bias $b=1$, the power spectrum in Equation (\ref{ClexLimber}) is just the mass power spectrum $P(k)$. If the selection function is too rapidly varying, one will need to keep additional terms in the Limber approximation. To illustrate this we take $W(z,z_0)$ to be a Gaussian centered at $z_0$ with variance $\sigma_z^2$ and compare the exact expression for $C_{gg}(\ell)$, Equation (\ref{Clexact}) with the Limber approximation to zeroth and second order in $1/\ell$ Equation (\ref{ClexLimber}). From the discussion in \ref{LimberDerivation} we expect the expansion to diverge for $\nu \lsim \nu_c \sim (r(z_0)/\sigma_r) $. For $\sigma_z=0.01$, this gives $\nu_c \approx 15$ at $z_0=0.3$ and $\nu_c \approx 30$ at $z_0=0.6$ . Comparison of the exact $C_\ell$ with the Limber expansion at different orders is shown in Fig. \ref{NarrowBinCompare}. For a given width, $\sigma$,  the Limber approximation is clearly more accurate for \tr{smaller} $z_0$ \tr{at a fixed $\ell$}. Very roughly, for $\ell > 5 \,r(z_0)/\sigma_r$ the $0^{th}$ order Limber approximation is accurate to $\sim 1\%$. 

\subsection{Cross-correlation of populations with small redshift overlap}
\begin{figure}
\begin{tabular}{cc}
\includegraphics[width=0.5\textwidth]{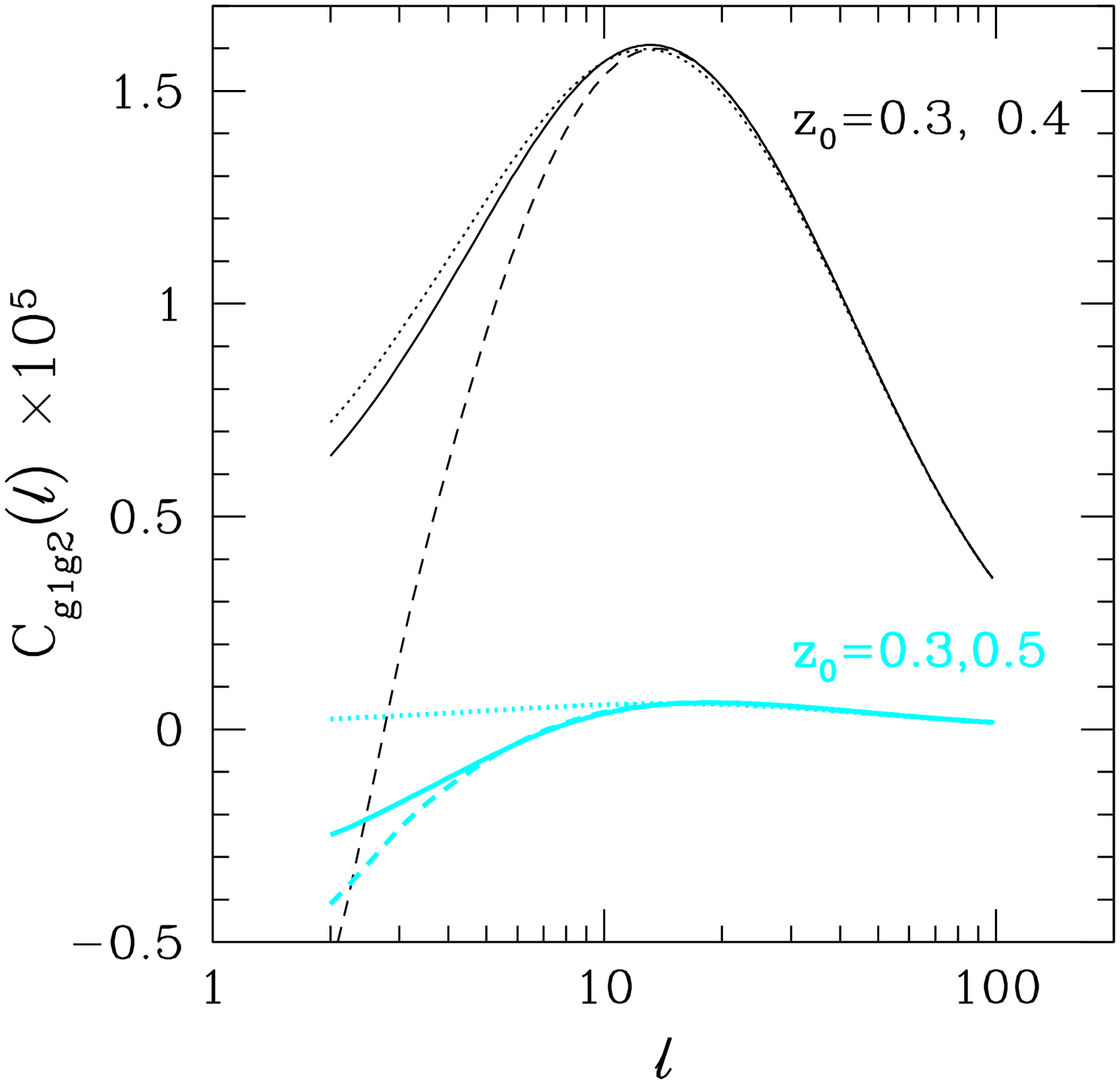}& \includegraphics[width=0.5\textwidth]{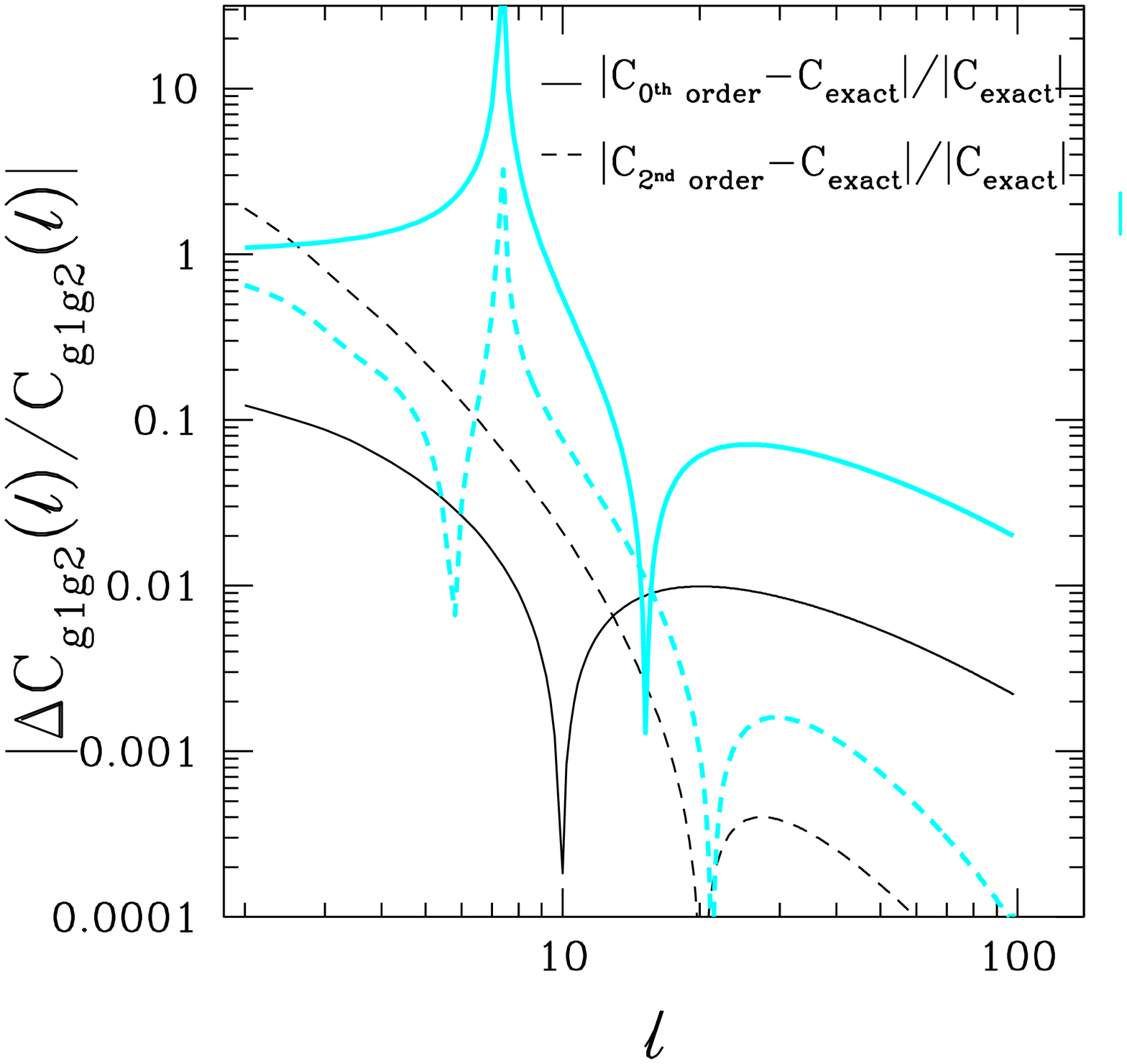}
\end{tabular}
\caption{Left panel: \tr{angular cross-power spectra between two samples with Gaussian width $\sigma_z=0.05$}. Upper black curves are for the cross power spectrum between bins at $z_0=0.3$ and $z_0=0.4$, lower cyan/gray curves are for more widely separated bins with $z_0=0.3$ and $z_0=0.5$. Solid lines show the exact power spectrum, dotted the $0^{th}$ order Limber \tr{formula} and dashed the Limber approximation to $2^{nd}$ order in $1/\nu$. Right panel: the difference between the curves shown \tr{on} left.  }
\label{CrossBins}
\end{figure}

Consider the cross-power spectrum between two source distributions with a small redshift overlap. Here, we \tr{will} use two selection functions with the same width but different mean redshifts. Cross-correlating different redshift bins is a tool for calibrating photometric redshifts (see, for example \cite{Schneider,Newman}). Distributions centered at different redshifts are also present in galaxy-lensing cross-correlation (which would more accurately correspond to a very broad and a narrow redshift distribution; for a review see \cite{Jain}). \tr{We then} use the expression given in Equation (\ref{littlefg}) for each sample, but allow the central redshifts $z_0$ to differ. Comparison of the Limber and exact calculation at different orders is shown in Figure \ref{CrossBins}\tr{.} As we had argued in \ref{LimberDerivation}, we see that the Limber approximation is less accurate for more widely separated redshift bins. \tr{Consequently, in this case, including the $2^{nd}$ order correction in Equation (\ref{ClexLimber}) could lead to a significant improvement in the accuracy of the Limber approximation}

\subsection{Cross-correlation of broad and narrow source distributions}
\begin{figure}
\begin{tabular}{cc}
\includegraphics[width=0.5\textwidth]{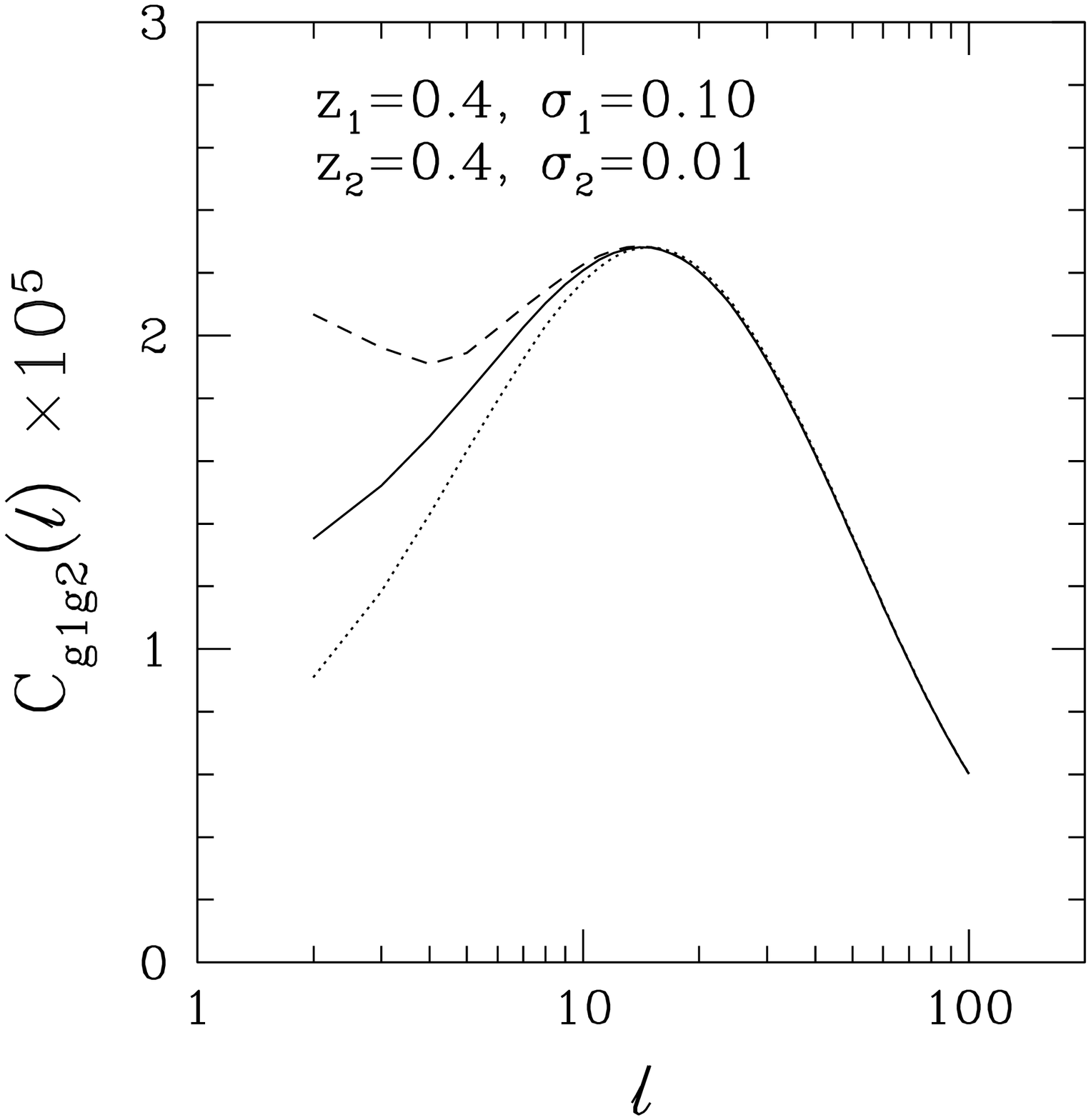}& \includegraphics[width=0.5\textwidth]{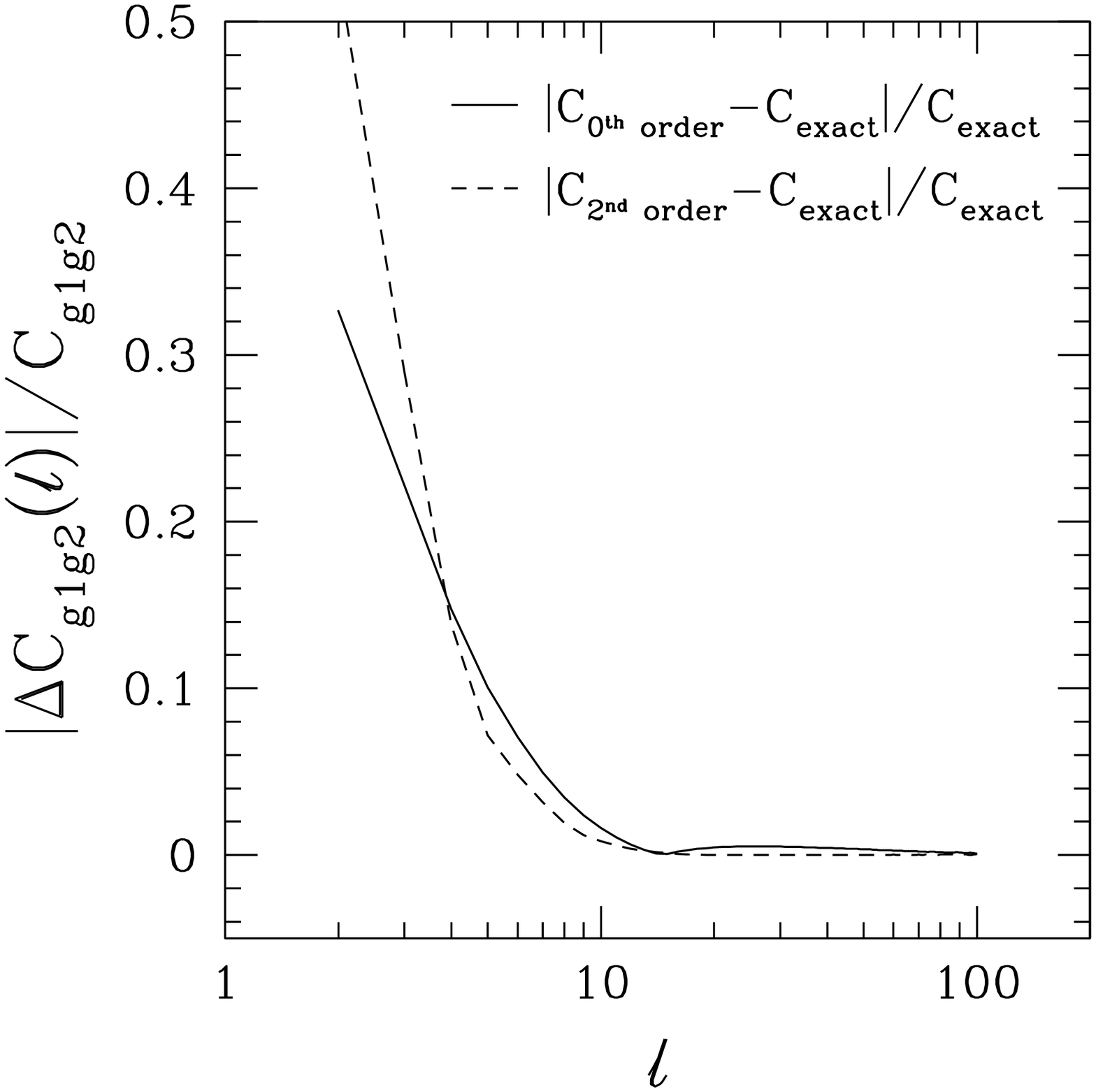}
\end{tabular}
\caption{Left panel: \tr{angular cross-power spectrum} between two redshift bins centered at $z=0.4$ with different widths $\sigma_1=0.10$ and $\sigma_2=0.01$. Solid lines show the exact power spectrum, dotted the $0^{th}$ order Limber calculation and dashed the Limber approximation to $2^{nd}$ order in $1/\nu$. Right panel: the difference between the curves shown at left.}
\label{Cross2Bins}
\end{figure}

Here we consider the cross-power spectrum between two sources with different redshift distributions, for example a broad and a narrow source distribution.  This is analogous to galaxy-lensing cross-correlation where the lensing weight function is broadly distributed and the galaxy selection function is narrow\tr{.} The limit that one source distribution becomes extremely broad is also analogous to the galaxy-CMB cross-correlation. We use the same Gaussian selection functions from the previous sections, but allow the widths of the two distributions to differ. This calculation is shown in Figure \ref{Cross2Bins}. Even though one redshift bin is narrow, since the other is broad the Limber approximation still works very well.

\section{Conclusions}
\label{Conclusion}
We have provided a series expansion for angular power spectra: The first term in this expansion gives the usual Limber approximation \cite{Limber,Kaiser,Kaiser2}, while the higher order terms are an extension to the approximation. The expression for the Limber approximation to second order in $1/\nu$ is given in Equation (\ref{ClexLimber}), higher order terms can be derived from Equations (\ref{Clexact}), (\ref{SeriesBessel}) and (\ref{LaplaceBessel}).  Figures \ref{NarrowBinCompare} through \ref{Cross2Bins} plot the accuracy of the Limber approximation at $0^{th}$ and $2^{nd}$ order in $1/\ell$ for a few examples. The Limber approximation is less accurate for rapidly varying projection kernels $f_A$ and $f_B$, or for $f_A$ and $f_B$ with small \tr{redshift} overlap.  The extended Limber approximation derived here can be applied to a variety of situations such as galaxy, weak lensing and CMB auto-correlations or to cross-correlations between the different projected distributions.

\tr{It is also worth pointing out that, even in the $0^{th}$ order Limber formula, replacing $\nu = \ell+1/2$ by $\ell$ (as is often done in the literature), will increase the error from ${\cal O}(\ell^{-2})$ to ${\cal O}(\ell^{-1})$. Therefore, simply using $\ell+1/2$, as obtained in our systematic derivation, can significantly improve the accuracy of the approximation.} 

\acknowledgements{
\tr{ML would like to thank Wenjuan Fang for helpful discussions. NA is supported by Perimeter Institute (PI) for Theoretical Physics.  Research at PI is supported by the Government of Canada through Industry Canada and by the Province of Ontario through the Ministry of Research \& Innovation. ML is supported by the DOE under DE-FG02-92ER40699, as well as the Initiatives in Science and Engineering Program at Columbia University.} ML is grateful for hospitality from the Harvard-Smithsonian Center for Astrophysics and Perimeter Institute where parts of this work were completed.}


\bibliographystyle{arxiv}
\bibliography{limber_bib}

\end{document}